\documentclass[twocolumn]{aastex631} 
\usepackage{relsize} 
\usepackage{newtxtext,newtxmath}
\usepackage{graphicx}	
\usepackage{amsmath}	
\usepackage[flushleft]{threeparttable}
\usepackage{changepage}
\usepackage{epstopdf}
\usepackage{amsopn,amsxtra,txfonts}
\usepackage{comment} 
\usepackage{longtable,threeparttablex}
\usepackage{enumitem}
\usepackage{fancyhdr}
\usepackage{booktabs}
\usepackage[hang,flushmargin]{footmisc}
\usepackage{gensymb}
\usepackage{tabularx}
\usepackage{ragged2e}
\usepackage{float}
\usepackage{enumitem}
\usepackage{hyperref}
\usepackage{wrapfig}
\usepackage{sidecap}
\usepackage{tcolorbox} 
\usepackage{xcolor} 
\usepackage{natbib}

\newcommand{\NiII}{\mbox{Ni\,{\sc ii}}}
\newcommand{\OI}{\mbox{O\,{\sc i}}}

\newcommand{\HI}{\mbox{H\,{\sc i}}}
\newcommand{\FeII}{\mbox{Fe\,{\sc ii}}}
\newcommand{\SII}{\mbox{S\,{\sc ii}}}
\newcommand{\SiII}{\mbox{Si\,{\sc ii}}}
\newcommand{\SiIII}{\mbox{Si\,{\sc iii}}}
\newcommand{\SiIV}{\mbox{Si\,{\sc iv}}}

\newcommand{\CIV}{\mbox{C\,{\sc iv}}}

\newcommand{\AlII}{\mbox{Al\,{\sc ii}}}

\newcommand{\msun}{\,$\rm M_{\odot}$}
 
\newcommand{\rholmc}{$\rho_{\rm LMC}$}

\newcommand{\kms}{\,km\,s$^{-1}$}
\newcommand{\vlsr}{$v_{\rm LSR}$}
\newcommand{\nsl}{28}
\defcitealias{Krishnarao2022}{K22} 

\newcommand{\oi}{\ion{O}{1}}
\newcommand{\feii}{\ion{Fe}{2}}
\newcommand{\siii}{\ion{Si}{2}}
\newcommand{\siiii}{\ion{Si}{3}}

\newcommand{\niii}{\ion{Ni}{2}}
\newcommand{\alii}{\ion{Al}{2}}
\newcommand{\sii}{\ion{S}{2}}

\begin{document}

\title{The Truncated Circumgalactic Medium of the Large Magellanic Cloud\footnote{Based on archival observations made with the NASA/ESA Hubble Space Telescope, obtained from the Data Archive at the Space Telescope Science Institute, which is operated by the Association of Universities for Research in Astronomy, Inc., under NASA contract NAS5-26555. Archival funding was associated with program 17053.}} 

\author[0000-0002-4157-5164]{Sapna Mishra}
\affiliation{Space Telescope Science Institute, 3700 San Martin Drive, Baltimore, MD 21218, USA}

\author[0000-0003-0724-4115]{Andrew J. Fox}
\affiliation{AURA for ESA, Space Telescope Science Institute, 3700 San Martin Drive, Baltimore, MD 21218}
\affiliation{Department of Physics \& Astronomy, Johns Hopkins University, 3400 N. Charles Street, Baltimore, MD 21218, USA}

\author[0000-0002-7955-7359]{Dhanesh Krishnarao}
\affiliation{Department of Physics, Colorado College , Colorado Springs, CO 80903, USA}

\author[0000-0001-9982-0241]{Scott Lucchini}
\affiliation{Center for Astrophysics | Harvard \& Smithsonian, 60 Garden Street, Cambridge, MA 02138, USA}
\affiliation{Department of Physics, University of Wisconsin- Madison, Madison, WI 53706, USA}

\author[0000-0003-2676-8344]{Elena D'Onghia}
\affiliation{Department of Physics, University of Wisconsin- Madison, Madison, WI 53706, USA}
\affiliation{Department of Astronomy, University of Wisconsin- Madison, Madison, WI 53706, USA}
\affiliation{INAF - Osservatorio Astrofisico di Torino, via Osservatorio 20, 10025 Pino Torinese (TO), Italy}

\author[0000-0003-4237-3553]{Frances H. Cashman}
\affiliation{Space Telescope Science Institute, 3700 San Martin Drive, Baltimore, MD 21218, USA}
\affiliation{Department of Physics, Presbyterian College, Clinton, SC 29325, USA}

\author[0000-0001-5817-0932]{Kathleen A. Barger}
\affiliation{Department of Physics \& Astronomy, Texas Christian University, Fort Worth, TX 76129, USA}

\author[0000-0001-9158-0829]{Nicolas Lehner}
\affiliation{Department of Physics and Astronomy, University of Notre Dame, Notre Dame, IN 46556, USA}

\author[0000-0002-7982-412X]{Jason Tumlinson}
\affiliation{Space Telescope Science Institute, 3700 San Martin Drive, Baltimore, MD 21218, USA}
\affiliation{Department of Physics \& Astronomy, Johns Hopkins University, 3400 N. Charles Street, Baltimore, MD 21218, USA}

\correspondingauthor{Sapna Mishra}
\email{smishra@stsci.edu}

\begin{abstract}
The Large Magellanic Cloud (LMC) is the nearest massive galaxy to the Milky Way. Its circumgalactic medium is complex and multi-phase, containing both stripped \HI\ structures like the Magellanic Stream and Bridge, and a diffuse warm corona seen in high-ion absorption. We analyze \nsl\ AGN sightlines passing within 35 kpc of the LMC with archival HST/COS spectra to characterize the cool ($T\approx10^4$\,K) gas in the LMC CGM, via new measurements of UV absorption in six low ions (\OI, \FeII, \SiII, \AlII, \SII, and \NiII) and one intermediate ion (\SiIII).
We show that a declining column-density profile is present in all seven ions, with
the low-ion profiles having a steeper slope than the high-ion profiles in \CIV\ and \SiIV\ reported by 
Krishnarao et al. 2022.
Crucially, absorption at the LMC systemic velocity is only detected (in all ions) out to 17 kpc. Beyond this distance, the gas has a lower velocity and is associated with the Magellanic Stream.
These results demonstrate that the LMC's CGM is composed of two distinct components:
a compact inner halo extending to 17 kpc, and a more extended stripped region associated 
with the Stream. The compactness and truncation of the LMC's inner CGM agree with recent simulations of ram-pressure stripping of the LMC by the Milky Way's extended corona.

\end{abstract}


\section{Introduction} \label{sec:intro}

The Large Magellanic Cloud (LMC), merely 50 kpc away \citep{Pietrzyski2013}, is the closest massive galaxy to the Milky Way (MW). Dynamic interactions with the MW and the Small Magellanic Cloud (SMC) are profoundly impacting the LMC and its gaseous circumgalactic medium (CGM). The gaseous structures in this system exhibit complex morphology and structure, as evidenced by the Magellanic Bridge, Magellanic Stream (hereafter the Stream), and Leading Arm (LA), which together with the LMC and SMC are known as the Magellanic System \citep[][see references therein]{Elena2016}. The Stream, an approximately 200$^{\degree}$-long tail of multi-phase gas, has been extensively mapped in both neutral hydrogen \citep{Putman2003, Bruns2005, Nidever2008, Nidever2010, Westmeier2018} and in ionized gas using UV surveys \citep{Richter2013, Fox2013, Fox2014} and H$\alpha$ surveys \citep{Putman2003a, Barger2017, BlandHawthorn2019}, see also \citet{Kim2024}.
The Stream is thought to have been stripped out of the Magellanic Clouds, either by tidal forces \citep{Fujimoto1977,Besla2012,Pardy2018} or ram pressure \citep[]{Moore1994, Diaz2011, Salem2015, Wang2019}.

In addition to the stripped gas, recent simulations and observations have revealed that the LMC is surrounded by its own diffuse ionized CGM, or ``corona''. Such a corona is motivated by the high mass of the LMC \citep[$>$10$^{11}$\msun;][]{Erkal2019,Petersen2021,Watkins2024} and the need to explain the high mass of ionized gas of the Stream,  which was previously not reproduced in tidal or ram-pressure models \citep{Lucchini2020, Lucchini2024}.
Observational evidence for the corona was provided by \citet{Krishnarao2022} -- hereafter \citetalias{Krishnarao2022}\citep[see also][]{deBoer1980, Wakker1998}. Using 28 HST/COS sightlines of background quasar passing through the LMC, \citetalias{Krishnarao2022} found a declining column-density profile of high ions (including \ion{O}{6}, \ion{C}{4} and \ion{Si}{4}) out to 35 kpc from the LMC. They conclude that the \ion{C}{4} and \ion{Si}{4} likely exist in the interfaces between the cooler ($\sim$10$^{4}$\,K) clouds and the hotter ($\sim$10$^{5.5}$\,K) corona traced directly by \ion{O}{6}. \citetalias{Krishnarao2022} found evidence for the cool LMC CGM, but did not discuss this phase in detail. Its properties and connection to the warm-hot corona are crucial for developing a complete picture of the LMC CGM and the history of the LMC. In this Letter, we present the results from a detailed study of the cool LMC CGM and its connection to the Magellanic System. 

\section{Sample and Analysis} \label{sec:sample_ana}

\begin{figure}[!hbt]
 \includegraphics[width=0.55\textwidth]{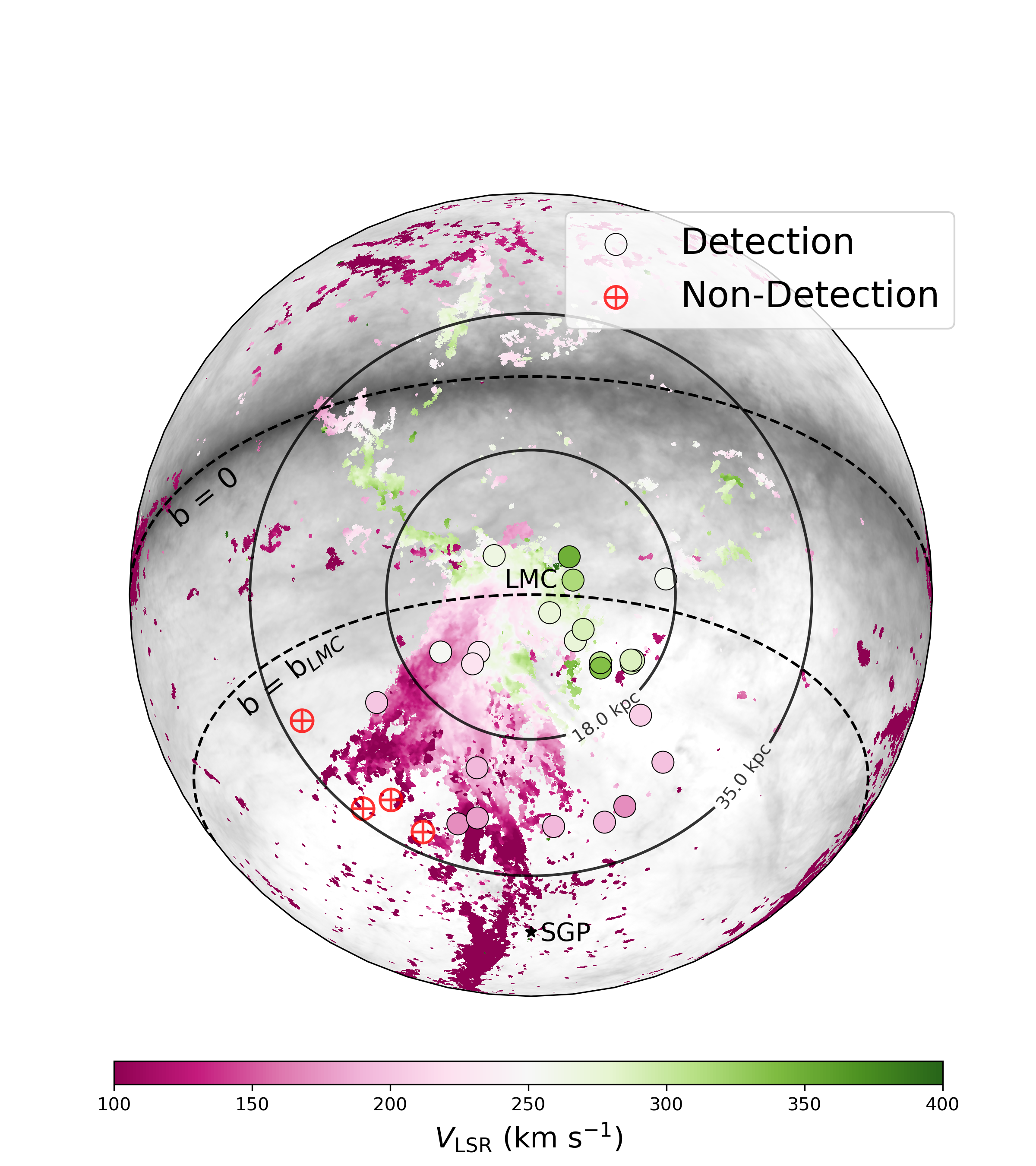}
 \caption{Velocity-coded map of the Magellanic System in an orthographic projection with LMC at the center, displaying the locations of 28 HST/COS background quasar sightlines (circles) from \citetalias{Krishnarao2022} with respect to the \HI\ emission (color-scale). The sightlines are color-coded based on the column-density-weighted central LSR velocity of Magellanic \SiII\ absorption. Magellanic absorbers are those with \vlsr\ $>$ 150\kms\ (see text in Section~\ref{sec:sample_ana}). Sightlines with \SiII\ non-detections at Magellanic velocities are marked with open red symbols. The map includes high-velocity 21~cm \HI\ emission data from \citet{Westmeier2018}, color-coded by LSR velocity.}
 \label{fig:lmc_sample} 
\end{figure}

We use the published archival dataset from \citetalias{Krishnarao2022} to map the CGM of the LMC. This dataset includes Hubble Space Telescope (HST)/Cosmic Origins Spectrograph (COS) spectra of 28 ultraviolet (UV)-bright background quasars that probe the LMC upto 35 kpc distance from the LMC's center. This distance corresponds to an angular separation of approximately 45$^{\degree}$. Sightlines beyond 45$^{\degree}$ are not included because at these large angles, geometrical distortions complicate the separation of absorption from the MW and the LMC. Fig.~\ref{fig:lmc_sample} shows the location of these sightlines (circles) relative to the LMC at the center. All the sightlines except two (RX\,J0503.1-6634 and PKS\,0552-640) probe the LMC CGM towards the Galactic south ($b<b_{\rm LMC}$). This is because high dust extinction from the MW hinders the identification of AGN closer to the galactic plane.

The data reduction steps for the COS spectra, including pipeline reduction, additional wavelength alignment, and removal of geocoronal airglow contamination in \ion{O}{1} $\lambda$1302 and \ion{Si}{2} $\lambda$1304, are based on the steps described in \citetalias{Krishnarao2022}. To merge the G130M and G160M grating COS spectra for each quasar, we join the spectra at the wavelength where the average SNR per pixel from both gratings is approximately equal. Each quasar spectrum is binned by 3 pixels using the {\tt Python} routine {\it SpectRes} \citep{Carnall2017}. To fit a global continuum across individual quasar spectra, we employ a method similar to that outlined in \citet{Mishra2022}. 

To trace the cool CGM of the LMC, we focus on the following transitions: \oi\ $\lambda$1302, \feii\ $\lambda\lambda$1143, 1144, 1608, \siii\ $\lambda\lambda$1190, 1193, 1260, 1304, 1526, \siiii\ $\lambda$1206, \alii\ $\lambda$1670, \sii\ $\lambda\lambda$1250, 1253, \niii\ $\lambda\lambda$1317, 1370. To identify the absorption associated with the Magellanic System, we impose a velocity threshold of \vlsr\ $>$ 150\kms, to avoid contamination from the Milky Way and therefore separate the Galactic and Magellanic components \citep{Boer1990,Wakker1997,Lehner2011, Richter2015}. For the maximum velocity threshold we used \vlsr\ $<$ 445\kms, which is calculated by finding the velocity dispersion around the LMC systematic velocity of 280\kms\ that encloses 99.9\% of Magellanic gas.

\begin{figure}[!hbt]
 \includegraphics[width=0.45\textwidth]{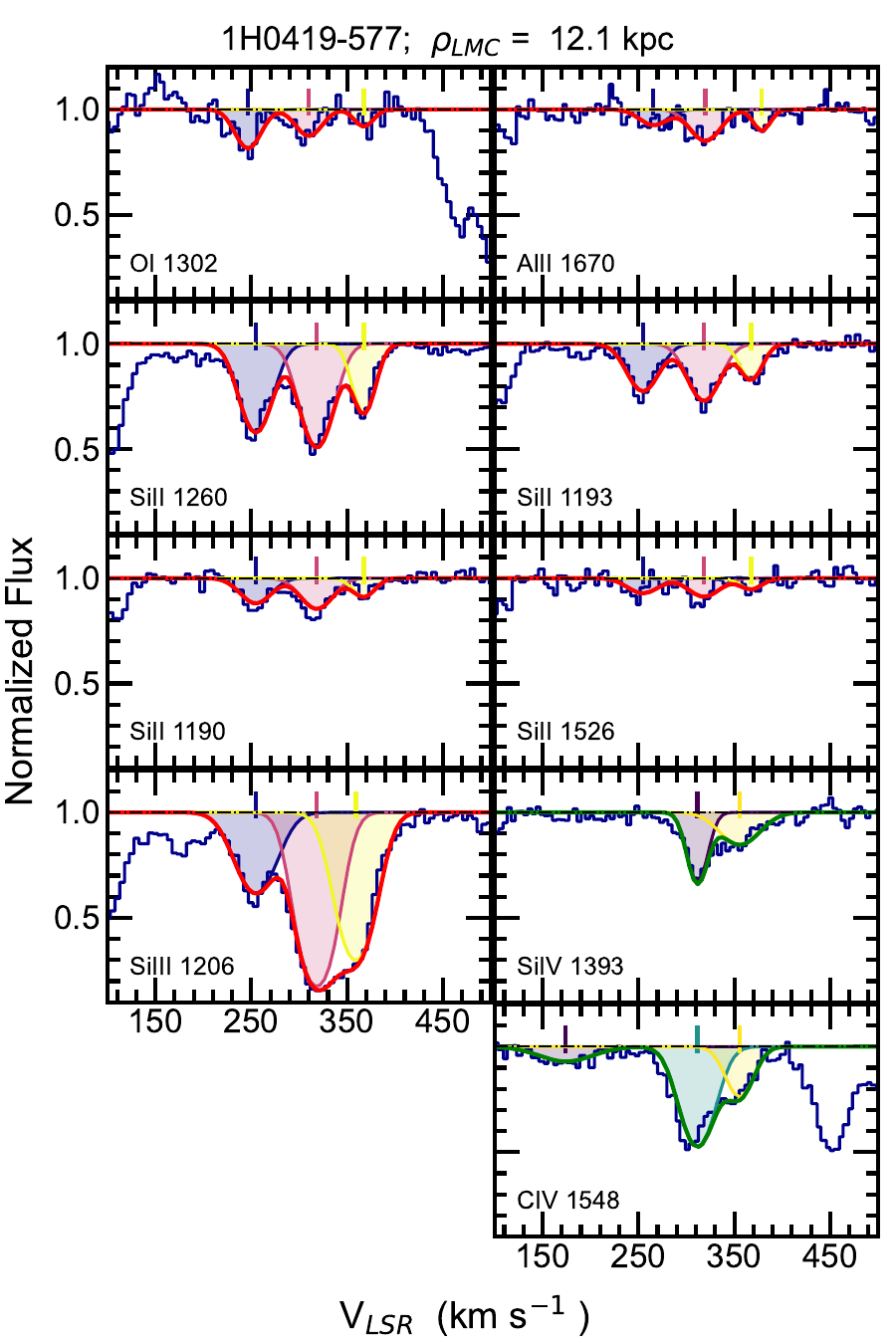}
 \caption{HST/COS metal-line profiles showing the absorption at Magellanic velocities toward 1H0419$-$577, as an example of our data quality and fitting methodology. Normalized flux is plotted against the LSR velocity in blue. The solid red lines represents the full Voigt profile fits to the low ions from this study, while the solid green lines shows the fits to the high ions from \citetalias{Krishnarao2022}. The individual components are shaded in different colors, with centroids marked by tick marks. The color scheme differs between low and high ions.}
 \label{fig:example_spec} 
\end{figure}

We use the Voigt-profile fitting software \texttt{VPFIT} (v12.3, \citealt{carswell2014}) to model these absorption profiles from the LMC CGM. Using a \(\chi^{2}\) minimization algorithm, \texttt{VPFIT} fits the line centroid, Doppler parameter ($b$), and column density ($N$) for each absorption component.  
To begin, we identify Magellanic components in each ion by visual inspection of the line profiles, and then compare across ions of the same ionization state (low or high) to establish the component structure in each sightline. We then simultaneously fit multiple lines of the same ion. For lines from different ions, we tie the centroid velocity of the components only when they are aligned based on the visual inspection.

For all ions, we label components as saturated absorption if the normalized flux drops below 0.2 for at least three consecutive rebinned pixels. For ions with multiple transitions, we exclude the saturated components from the fitting process unless there are unsaturated components available from the weaker transitions. For ions with only one transition, we  provide a lower limit for the column density if there is even a single saturated component (commonly the case for \ion{Si}{3}). If there are velocity offsets between the transitions of the same ion falling on the G130M and G160M gratings, we applied a shift to align these transitions together.  
For non-detected transitions of an ion, we measure the 3$\sigma$ upper limit using the SNR of the absorption-free region and assume the line extends over a width equal to the mean $b$-value of detected transitions for that ion. The adopted upper limit for an ion with multiple transitions is taken from the upper limit from the strongest transition (with the highest oscillator strength), since this gives the strongest constraint.

The absorption parameters for the high ions \ion{C}{4}\ and \ion{Si}{4} are taken from \citetalias{Krishnarao2022}. However, for the sightlines RBS\,567 and IRAS\,Z06229-6434, we identify several additional Magellanic components, specifically two \ion{C}{4}\ components in each sightline, and two and one \ion{Si}{4}\ components toward RBS\,567 and IRAS\,Z06229-6434, respectively. We fit these new components in a similar manner to the low ions using \texttt{VPFIT}.

An example of our Voigt fits to the UV absorption-line profiles is shown in Fig.~\ref{fig:example_spec}, for the sightline towards 1H0419$-$577, which passes 12.1 kpc from the LMC. These spectra show multiple Magellanic components in both the low and high ions. The relative strength of the low and high ions varies between components. The stack plots for the remaining sightlines are shown in Figs.~\ref{fig:appendix_set1} $-$ \ref{fig:appendix_set6}. In Table~\ref{tab:vpfit_param}, we present the Voigt-fit parameters for all Magellanic absorption components (\vlsr$>$150\kms) in the sample.

\section{Results}
\label{sec:results}

\begin{figure*}[!hbt]
 \includegraphics[width=1.05\textwidth]{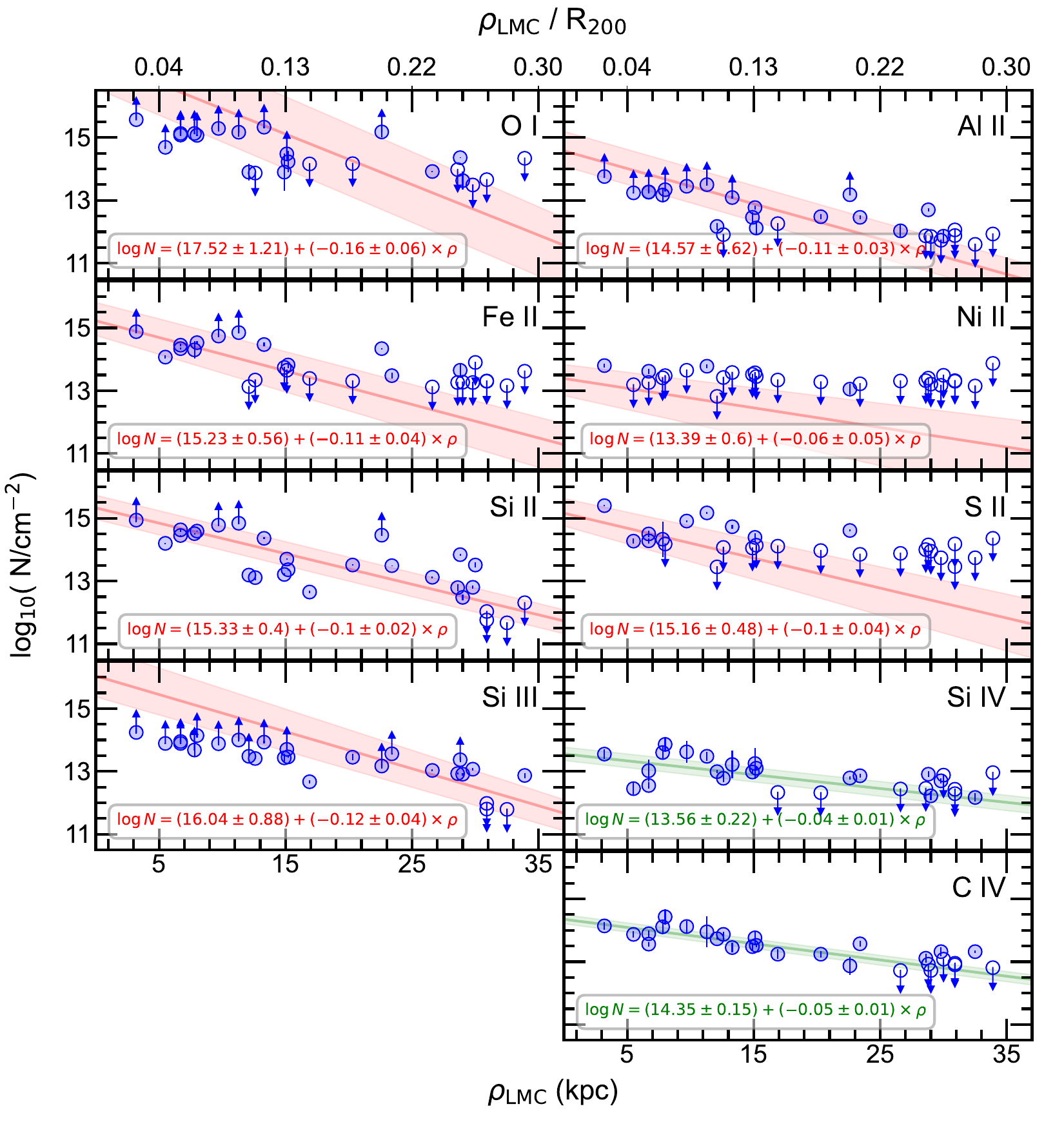}
 \caption{Total column density of Magellanic gas as a function of LMC impact parameter. The profiles of six low ions and one intermediate ion (\SiIII) are shown in red. For comparison, the profiles for the high ions \ion{Si}{4} and \ion{C}{4} are included in green; these data are taken from \citetalias{Krishnarao2022} along with additional \ion{C}{4} and \ion{Si}{4} components identified towards RBS\,567 and IRAS\,Z06229-6434. Open circles with downward arrows indicate $3\sigma$ upper limits, while circles with upward arrows represent lower limits from saturated lines. The 
 best-fit log-linear relations for MCMC runs with $1\sigma$ scatter (shaded regions) are shown as solid lines with the relation labeled in the bottom-left corner of each panel. The best-fit relations using the normalized impact parameter instead of $\rho_{\rm LMC}$ are the same with the slopes scaled by R$_{\rm 200}$ = 115$\pm$15 kpc \citepalias[see][]{Krishnarao2022}. In this plot, Magellanic components are defined as those with \vlsr$>$150\kms.}
 \label{fig:res_radial_profile} 
\end{figure*}

We present in Fig.~\ref{fig:res_radial_profile}, the total column density of Magellanic components (\vlsr $>$ 150\kms) for six low ions (\OI, \AlII, \FeII, \NiII, \SiII,   \SII) and one intermediate ion (\SiIII) plotted against both the impact parameter (\rholmc, bottom x-axis) and the normalized impact parameter (\rholmc ~/ R$_{\rm 200}$\footnote{R$_{\rm 200}$ is the radius enclosing a mean overdensity of 200 times the critical density.}, top x-axis). We use R$_{\rm 200}$ = 115$\pm$15 kpc for LMC \citepalias[see][]{Krishnarao2022}. For comparison, we have included the column density measurements for \ion{C}{4}\ and \ion{Si}{4}\ from \citetalias{Krishnarao2022} along with our new column density estimates for these ions along the RBS\,567 and IRAS\,Z06229-6434 sightlines.  A declining column density profile is observed in all nine ions shown.
\begin{figure*}[!hbt]
 \includegraphics[width=0.99\textwidth]{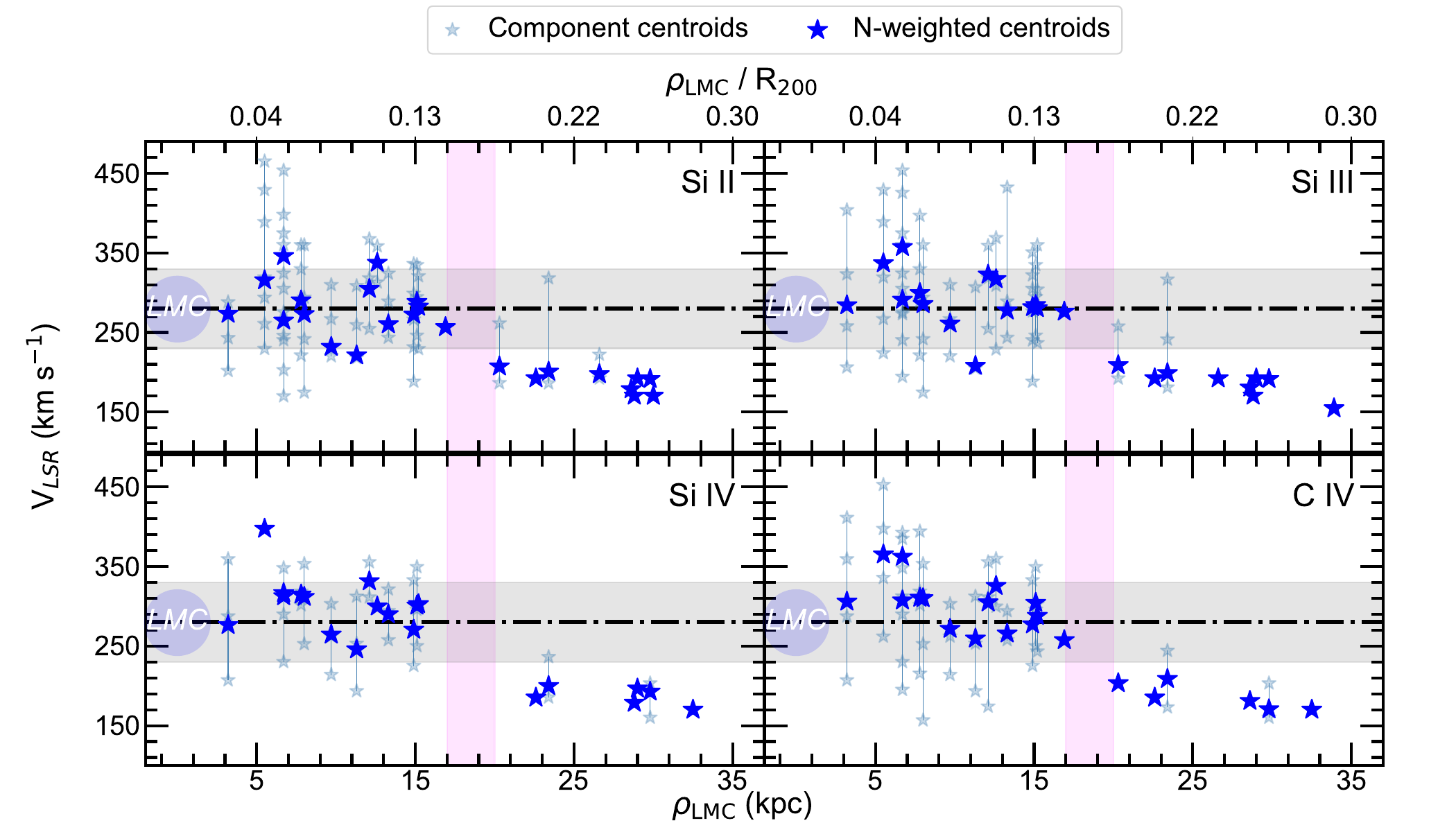}
 \caption{Velocity-position analysis of gas in the LMC CGM. LSR velocity is plotted against LMC impact parameter both for individual components (steel-blue stars) and for the column density-weighted central velocity of all Magellanic components along each sightline (large blue stars). The solid vertical lines in steel-blue connect the different velocity components of each sightline and illustrate the dispersion of LMC gas in each direction. The horizontal dashed black line represents the LMC systemic velocity of 280\kms.  The shaded gray area indicates the region within $\pm 50$\kms\ of the LMC velocity. The shaded area in magenta marks the region where the gas velocities change substantially; this truncation point separates the inner CGM from the Stream.}
 \label{fig:res_velocity_LMC} 
\end{figure*}

To model our dataset, we use Bayesian regression with censoring via the \texttt{PyMC3} package \citep{Salvatier2016}. This approach effectively accounts for upper limits and lower limits in our data. \texttt{PyMC3} is a Python-based probabilistic programming framework, that employs Markov chain Monte Carlo (MCMC) and variational inference algorithms to estimate the posterior distributions of model parameters. A detailed illustration of this method is described in \citet{Zheng2024} (see their Appendix C). For each ion profile, we model the column density (log $N$) and impact parameter (\rholmc) simultaneously for detections, lower limits from saturated absorption, and upper limits from non-detections using a log-linear relation (exponential fit), mathematically expressed as  log ($N$/cm$^{-2}$) = c + m ($\rho_{\rm LMC}$/kpc). A declining trend in the total column density with increasing \rholmc\ for all the ions is evident in Fig.~\ref{fig:res_radial_profile}. The slopes of the log-linear relation for the low and intermediate ions are consistent with each other within 1$\sigma$. Additionally, the column density profiles for low and intermediate ions are steeper compared to those for high ions, as indicated by the slopes of the fits, with the only exception being the \NiII\ profiles. However, since \NiII\ is detected along only four quasar sightlines, the fit has higher uncertainty, and hence we do not infer anything conclusive from the \NiII\ profile.

Next, we explore the LMC CGM kinematics using the measured velocity centroids of low and high ions. In Fig.~\ref{fig:res_velocity_LMC}, we show the centroid velocities of individual components with \vlsr$>$150\kms\ as a function of \rholmc\ for \SiII, \SiIII, \ion{Si}{4}, and \ion{C}{4}. These four ions are chosen to represent gas in different phases, although the results are consistent with all nine ions presented in Fig.~\ref{fig:res_radial_profile}. We observe more kinematic complexity, indicated by more velocity components, closer to the LMC (with an average of $\sim$3 components per sightline for \rholmc\ $<$17 kpc) compared to farther regions (with an average of $\sim$1 component per sightline for \rholmc\  $>$17kpc). This result holds separately for low, intermediate, and high ions. The dispersions in the velocity centroids for $<$17 kpc ($>$17 kpc) are 67\kms\ (42 \kms) for low, 65\kms\ (43 \kms) for intermediate, and 67\kms\ (42 \kms) for high ions respectively. The mean and standard deviation of centroid velocities of individual LMC CGM components in \SiII, \SiIII, \ion{Si}{4}, and \ion{C}{4} are 276$\pm$71, 281$\pm$71, 275$\pm$60, and 281$\pm$72\kms\ respectively, all closely consistent with each other.

We then estimate the column density-weighted velocity centroids, defined as \( v_{\rm weighted} $=$\frac{\mathlarger{\Sigma}(v_{\rm comp} \times {\rm N_{comp}})}{\mathlarger{\Sigma}~{\rm N_{comp}}} \), for each sightline, to give an estimate of the {\it mass-weighted average velocity} of the Magellanic gas.  In Fig.~\ref{fig:res_velocity_LMC}, we plot these weighted velocities for the ions as a function of \rholmc. The mean and standard deviation of the weighted velocity for the four ions are consistent with each other, with an average value of 274$\pm$56\kms. In Fig.~\ref{fig:res_velocity_LMC}, the horizontal black dashed line indicates the systematic velocity of the LMC of 280\kms, with the shaded gray region representing the $\pm$50~\kms\ region around it, consistent with our average dispersion of 56\kms\ for the LMC in all phases. From the figure, we observe a drop near 17 kpc 
(shown as a shaded region in magenta) where the $N$-weighted centroids fall sharply for all ions, with the gas further out only observed at lower LSR velocities. We confirm that the velocity distributions of absorbers inside and outside 17 kpc differ significantly at the 99.9\% confidence level using a two-sample Kolmogorov–Smirnov test. We argue in Section~\ref{sec:discussion} that the gas beyond 17 kpc mostly traces the stripped gas in the Stream. Therefore, in the next section, we use a more restrictive limit of $v>$230\kms\ to identify the gas with the (non-stripped) LMC CGM.

\section{Discussion}
\label{sec:discussion}

The relationship between metal-line column density and impact parameter has been widely used to characterize the CGM of galaxies \citep[e.g.][]{Prochaska2011, Tumlinson2013}. Fig.~\ref{fig:res_radial_profile} shows this relationship for six low, one intermediate, and two high ions in the LMC CGM. We find a declining trend in these profiles as inferred from the slope of the log-linear fits given in each panel. The declining column density profile is a characteristic signature of the CGM (\citealt{Tumlinson2017}). We note that all our sightlines probe at least 3$^{\degree}$ (i.e. $\gtrsim$ 3 kpc) off the LMC, which is beyond the known extent of LMC winds arising from stellar feedback \citep[see][]{Barger2016, Zheng2024b}. 

From the slopes of the LMC column density profiles (Fig.~\ref{fig:res_radial_profile}), 
the following observations can be made: 
(1) All the low-ion and intermediate-ion profiles have similar slopes, consistent within the errors. 
(2) The high-ion slopes from \SiIV\ and \CIV\ are consistent with each other. 
(3) The low-ion slopes are steeper by a factor of two than the high-ion slopes \citep[similar to the M31 CGM;][]{Lehner2020}. 
The declining column density profiles with \rholmc\ are seen for \emph{all} ions
at 3$\sigma$ significance (see Fig.~\ref{fig:res_radial_profile}).\par

One interpretation of the declining column density profiles across multiple ions is that the gas originates from the multiphase CGM of the LMC. The steeper slope of the low ions compared to the high ions indicates that the LMC CGM is (relatively) more ionized at larger distances. One possible multiphase CGM model is the interface or boundary-layer scenario as favored by \citetalias{Krishnarao2022}, where high ions, such as \SiIV\ and \CIV, arise at the turbulent or conductive interfaces between 10$^{4}$\,K low-ion clouds and the hot $\approx$10$^{5.5}$\,K diffuse gas of the Magellanic Corona \citep[e.g.][]{Kwak2015, Claude2023}. This interface scenario is consistent with our kinematic findings shown in Fig.~\ref{fig:res_velocity_LMC}, which show that the low and high ions are kinematically related. On the other hand, the flatter profiles of high ions compared to low ions may also result if the high-ion gas includes components produced by both photoionization and collisional ionization. However, \citetalias{Krishnarao2022} found that the influence of photoionization is strongest within 7 kpc of the LMC, while beyond that distance, the gas traced by \SiIV\ and \CIV\ is almost entirely collisionally ionized, making this explanation less likely.

\begin{figure*}[!hbt]
 \includegraphics[width=0.99\textwidth]{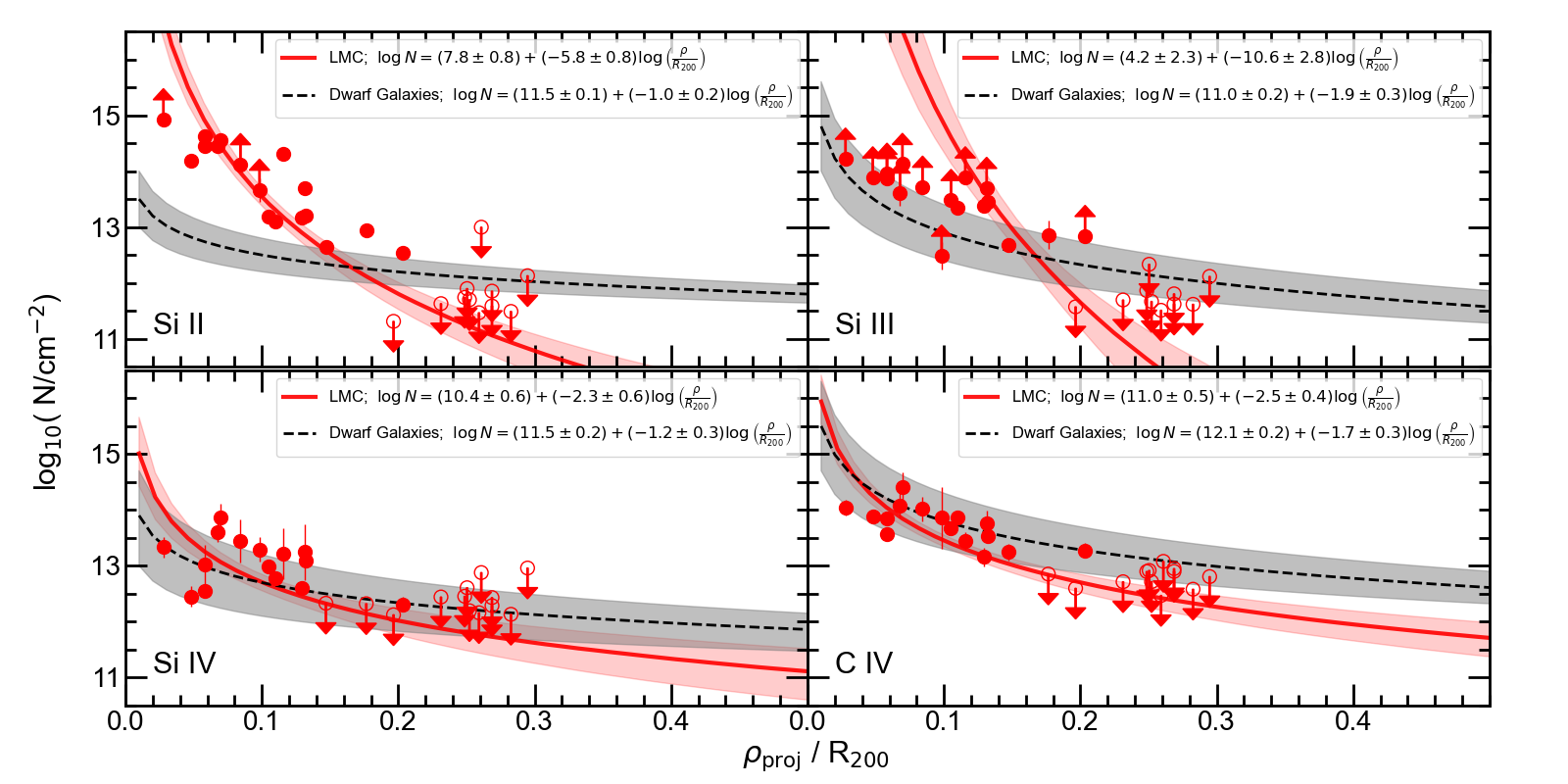}
 \caption{Column density profiles of the CGM in four low ions, comparing the LMC to isolated dwarf galaxies in the literature. The red points show the total column density of LMC components at \vlsr $>$ 230\kms\ (a range chosen to exclude the Stream) as a function of impact parameter normalized by the virial radius. Open symbols indicate the 3$\sigma$ upper limits for non-detections. The 50th percentile solution of the power-law model for MCMC runs for the LMC is shown in red solid line with 1$\sigma$ scatter of the fit shown in shaded red. For comparison, the profiles from the CGM of isolated dwarf galaxies taken from \citet{Zheng2024} are shown in black in each panel with 1$\sigma$ uncertainty in gray. 
 }
 \label{fig:res_radial_profile_norm} 
\end{figure*}

In our kinematic analysis (Fig~\ref{fig:res_velocity_LMC}), the weighted centroid velocities drop sharply beyond 17 kpc for both low-ions and high-ions, with absorbers beyond 17 kpc predominantly found at LSR velocities $<$230\kms. We confirm that this decrease is neither due to the reduced number of sightlines beyond 17 kpc (see Fig.~\ref{fig:lmc_sample}) nor to a decrease in the sensitivity of the spectra beyond 17 kpc. These lower-velocity absorbers are associated with the Magellanic Stream, as seen in Fig.~\ref{fig:lmc_sample}. Sightlines within 17 kpc of the LMC show \SiII\ at the LMC's systemic velocity of 280\kms. Beyond 17 kpc, the \SiII\ absorbers have much lower velocities consistent with the Stream. To explore the role of the Stream in more detail, we examined the low- and high-ion column densities as a function of impact parameter for absorbers with velocities between 150 and 230\kms\ (i.e. Stream velocities). These profiles do not show a significant declining trend with \rholmc. In addition, we find no correlation between the column density and \rholmc\ for these lower-velocity absorbers using the Spearman rank correlation test, further indicating that the gas beyond 17 kpc is associated with the Stream rather than the LMC.

In principle, the compactness of a galaxy's CGM can be explained by: (i) the galaxy not being massive enough to retain a larger CGM that extends up to its virial radius \citep[see][]{Bordoloi2014}, and/or (ii) the CGM being truncated due to strong interactions with the surrounding environment. Since it is now well-established that the LMC is an intermediate-mass galaxy with $M_{\rm halo}>10^{11}$\msun\ \citep{Erkal2019, Watkins2024}, the first scenario is unlikely, leaving environmental effects as the likely explanation for the compactness of the LMC CGM. 

To explore the environmental effects, we compare the CGM of the LMC with that of isolated dwarf galaxies using the recent sample of \citet{Zheng2024}, which focuses on the metal content in the CGM of nearby dwarf galaxies. This sample consists of 45 isolated dwarf galaxies with 56 galaxy-quasar pairs having $z$ = $0.0-0.3$ and mass range of $M_{*} \approx 10^{6.5-9.5}$ \msun, including the COS-Dwarfs sample \citep{Bordoloi2014}. For the LMC CGM, we do not impose any impact parameter cut to ensure an unbiased comparison with the isolated dwarf galaxy sample, but we only include absorbers with \vlsr $>$230 \kms\ to exclude Stream contributions. For sightlines beyond 17 kpc, we treat them as non-detections and estimate upper limits at the LMC's systematic velocity of 280 km/s. In Fig.~\ref{fig:res_radial_profile_norm}, we show the total column density of\ SiII, \SiIII, \SiIV, and \CIV\ as a function of normalized impact parameter. To compare the corresponding column density profiles for the isolated dwarf galaxies, we used the best-fit log-log relation (power-law fit) presented in Figure 4 of \citet{Zheng2024}, which has the form $\log N$=$\log N_{0}$ + $k \log \frac{(\rho}{R_{\rm 200}})$. 
We see clearly that the LMC profiles are steeper than the profiles of the isolated dwarf galaxies. This difference is strongest for the low-ion \ion{Si}{2}, also strong for the intermediate-ion \ion{Si}{3}, and still present but weaker for the high-ion \ion{Si}{4}. The distinct differences in the profile of the LMC halo compared to other dwarf galaxies can be attributed to the strong dynamical interactions of the LMC with the MW and SMC, i.e. to the highly non-isolated nature of the LMC. Therefore, our observations of a steep column density profile for the LMC CGM, despite its higher stellar mass and halo mass than the galaxies in the \citet{Zheng2024} sample, suggest that strong environmental effects are disturbing and stripping the LMC CGM.

The 17 kpc truncation radius of the LMC CGM can be understood as a ram-pressure effect. A recent simulation by \citet{Zhu2024} estimates the survival of an LMC-like CGM against ram pressure stripping in a MW-type environment (see their Section~5.4). By employing an isothermal spherical CGM with a power-law density profile (power-law index of -2) and using pericentric ram pressure value of $P_{\rm ram} \approx 2 \times 10^{-13}$ dyne cm$^{-2}$ \citep{Salem2015}, the authors predict a maximum CGM stripping radius of 15 kpc, which aligns well with our observed LMC truncation radius of 17 kpc. The authors found only 10\% of the initial CGM mass survives. It is noteworthy that the LMC's CGM may no longer be spherical and could be compressed towards the front (i.e north-side with $b>b_{\rm LMC}$), as evident from the generation of bow shocks \citep{Setton2023}. Nonetheless, our observed LMC CGM truncation radius of 17 kpc is also in reasonable agreement with the simulations of \citet{Lucchini2020}, \citet{Lucchini2024}, and \citet{Carr2024}, which account for this front-side compression. Specifically, the LMC corona in \citet{Lucchini2024} extends $\approx$20$^{\degree}$  to either side of the LMC, corresponding to $\sim$18 kpc at the LMC distance. However, this region towards the north ($b>b_{\rm LMC}$) currently lacks UV sightlines (see Fig. 1) and therefore remains unprobed observationally. We have an approved HST/COS Cycle 32 program to probe the LMC CGM in this region (PI: S. Mishra).

\section{Summary} \label{sec:summary}

We have presented new findings on the LMC CGM using HST/COS G130M and G160M spectra of 28 quasar sightlines extending up to 35 kpc from the LMC. We trace the cool ($\sim$10$^{4}$ K) gas using six low ions: \OI, \AlII, \FeII, \NiII, \SiII, \SII,  and one intermediate ion: \SiIII. Our results are supplemented by high-ion data (\SiIV\ and \CIV) from  \citetalias{Krishnarao2022}. We observe a declining column density profile for both low-ion and high-ion column densities as a function of the impact parameter, with the low ions declining more steeply than the high ions. The kinematic structure of both the cool and warm phases becomes more complex closer to the LMC compared to larger impact parameters.\par 
Importantly, we observe a break in the LMC CGM properties at 17 kpc. Inside 17 kpc, the CGM absorption is found within $\approx$50\kms\ of the LMC systemic velocity of 280\kms. Beyond 17 kpc, the absorption is predominantly found at the much lower velocities ($<$230\kms) of the Stream, indicating a clear truncation in the LMC CGM at this distance. The truncation radius is in good agreement with recent simulations \citep{Zhu2024, Lucchini2020, Lucchini2024, Carr2024}. Our finding of a truncated LMC CGM supports the picture of a high-mass LMC on its first infall passage that has lost most (but not all) of its CGM  to ram pressure stripping by the MW halo. As a result, the MW halo has gained mass, but the LMC halo has still survived. The survival of a small halo is important for the LMC's evolution,  as the halo protects its interstellar gas from being stripped and allows star formation in the LMC to continue.

\vspace{0.4 cm}
{\it Acknowledgments:}
We thank the referees for their constructive and helpful reports. Support for program 17053 was provided by NASA through a grant from the Space Telescope Science Institute, which is operated by the Association of Universities for Research in Astronomy, Inc., under NASA contract NAS5-26555. 

\appendix

\setcounter{table}{0}
\setcounter{figure}{0}
\renewcommand{\thefigure}{A\arabic{figure}}
\renewcommand{\thetable}{A\arabic{table}}  

\begin{table*}[!hbt]
\small
\caption{Absorption parameters of Magellanic components derived from Voigt-profile fitting}
\label{tab:vpfit_param} 
\begin{tabular}{lllllllllll}
    \hline
\multicolumn{1}{c}{Quasar Name}  &  GLON    &   GLAT   &   $\rho$  &   Ion  &    $v_{\rm LSR}$   &      $b$        &    log\,$N$     &      Reference \\
                                 &   (deg)  &  (deg)   &    (kpc)  &        &   (km s$^{-1}$)    & (km s$^{-1}$)   &  (cm$^{-2}$)    &                 \\
  (1)                            &   (2)    &    (3)   &   (4)     &  (5)   &    (6)             &    (7)          &    (8)          &         (9)      \\
\hline 
RX J0503.1$-$6634           &  277.18   & $-$35.42   & 3.2   &  \ion{O}{1}     &  229.6$\pm$15.1 &   46.8$\pm$1.6   &   $>$15.18        &      This work \\
                            &            &           &       &  \ion{O}{1}     &  296.6$\pm$13.8 &   51.0$\pm$2.5   &   $>$15.34        &      This work \\
                            &            &           &       &  \ion{O}{1}     &  374.4$\pm$2.3  &   10.1$\pm$3.6   &   13.54$\pm$0.19  &      This work \\
                            &            &           &       &  \ion{Al}{2}    &  162.6$\pm$3.2  &   36.7$\pm$56.2  &   12.34$\pm$0.57  &      This work \\
                            &            &           &       &  \ion{Al}{2}    &  204.7$\pm$2.8  &   10.7$\pm$4.7   &   12.31$\pm$0.27  &      This work \\
--                          &    --      &   --      &  --   &   --            &    --           &     --           &     --            &         --     \\
\hline
\end{tabular}

\begin{tablenotes}\small
\item Notes --  (1) Quasar name; (2)-(3) galactic longitude and latitude; (4) impact parameter; (5) name of the ion; (6)-(8) centroid velocity, line width, and column density of absorption component from VPFIT; (9) reference for the measurements. The full table is available in the online version. Only a portion is presented here to illustrate its format and content.
\end{tablenotes}
\end{table*}

\begin{figure*}[!hbt]
 \includegraphics[width=0.95\textwidth]{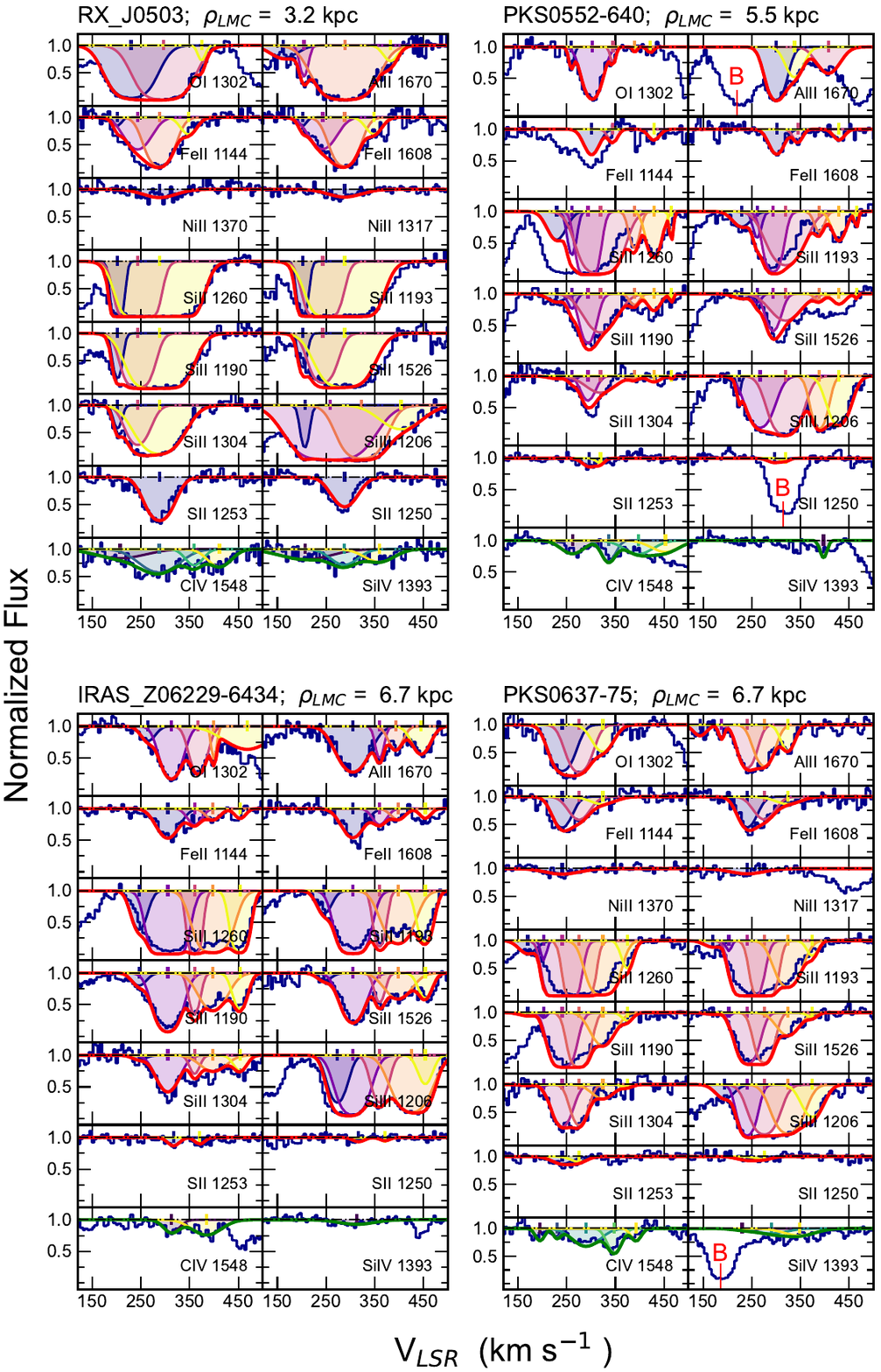}
 \caption{Same as Fig.~\ref{fig:example_spec}. We indicate the locations of blended regions with 'B' in red and exclude these regions from the fitting.}
 \label{fig:appendix_set1} 
\end{figure*}

\begin{figure*}[!hbt]
 \includegraphics[width=0.95\textwidth]{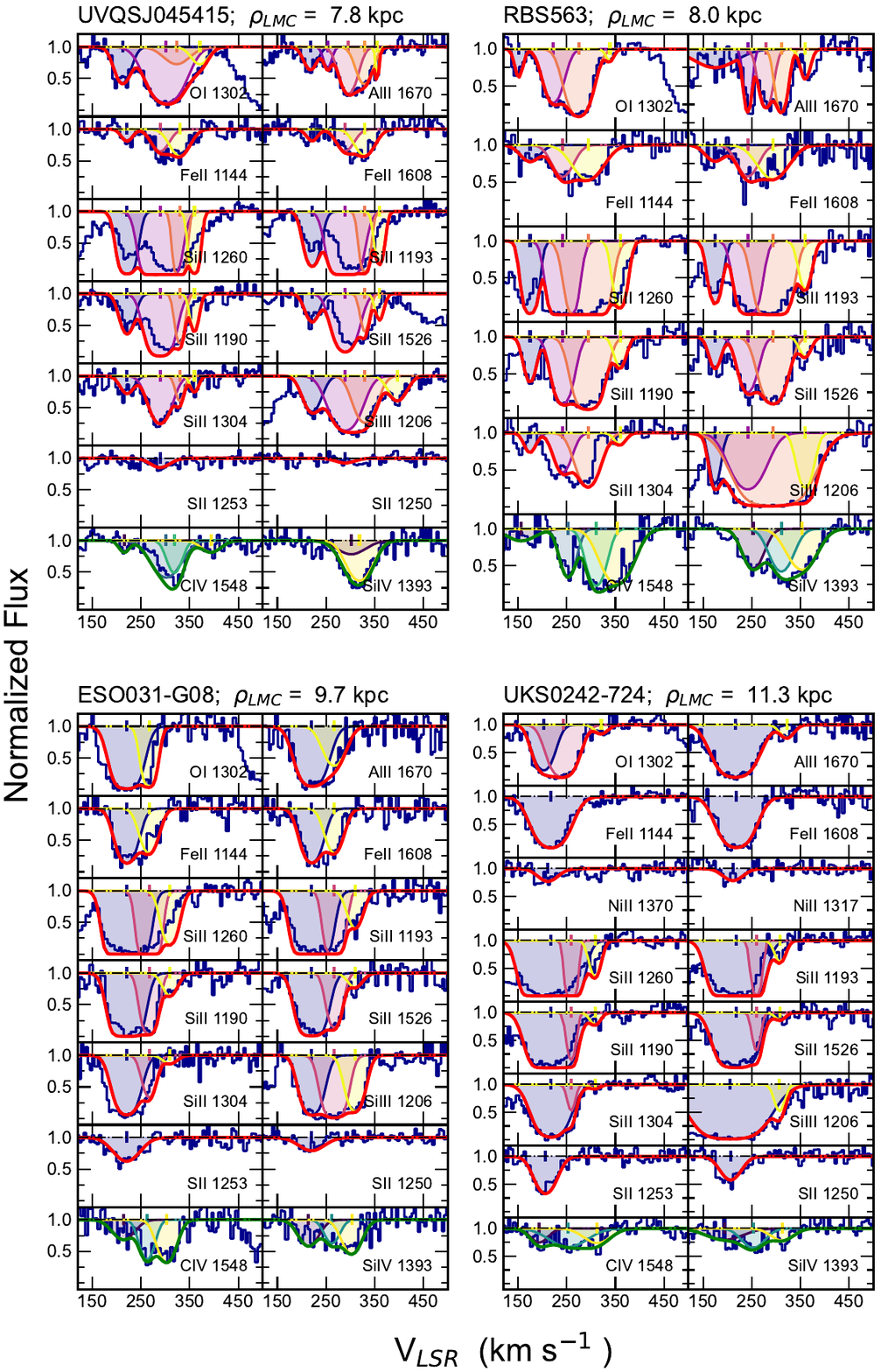}
 \caption{Same as Fig.~\ref{fig:example_spec}.}
 \label{fig:appendix_set2} 
\end{figure*}

\begin{figure*}[!hbt]
 \includegraphics[width=0.95\textwidth]{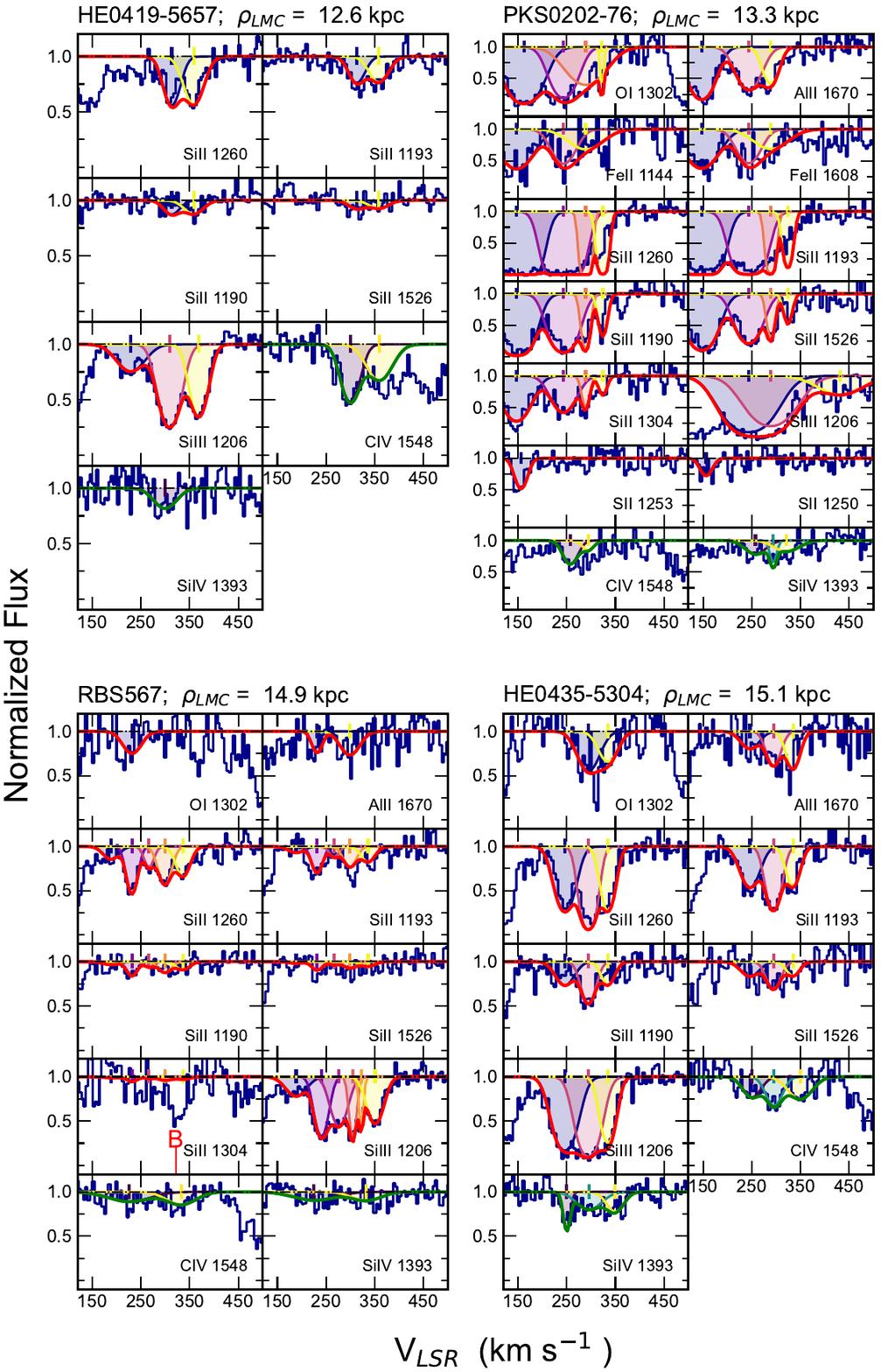}
 \caption{Same as Fig.~\ref{fig:example_spec}.}
 \label{fig:appendix_set3} 
\end{figure*}

\begin{figure*}[!hbt]
 \includegraphics[width=0.95\textwidth]{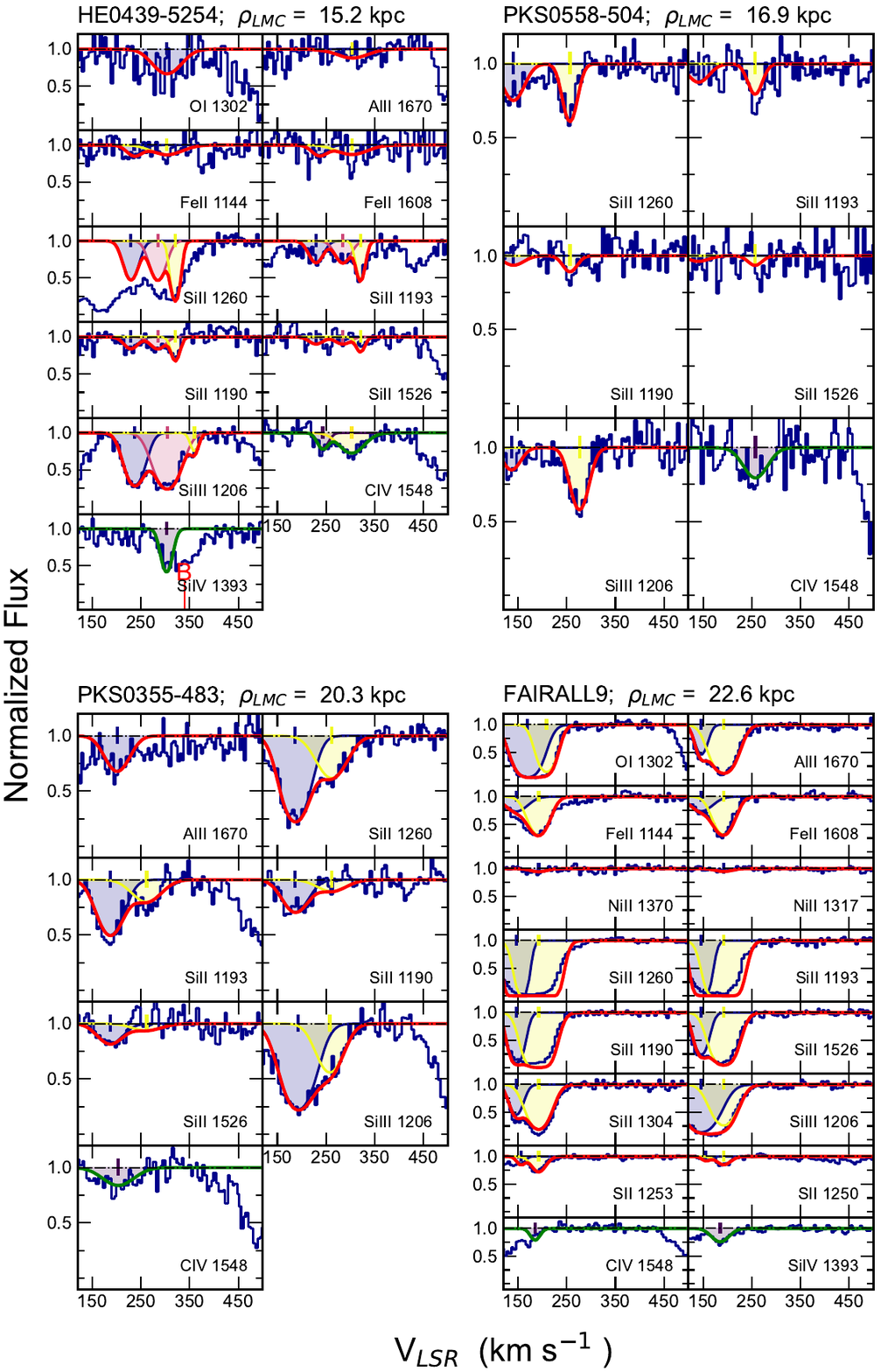}
 \caption{Same as Fig.~\ref{fig:example_spec}.}
 \label{fig:appendix_set4}
\end{figure*}

\begin{figure*}[!hbt]
 \includegraphics[width=0.95\textwidth]{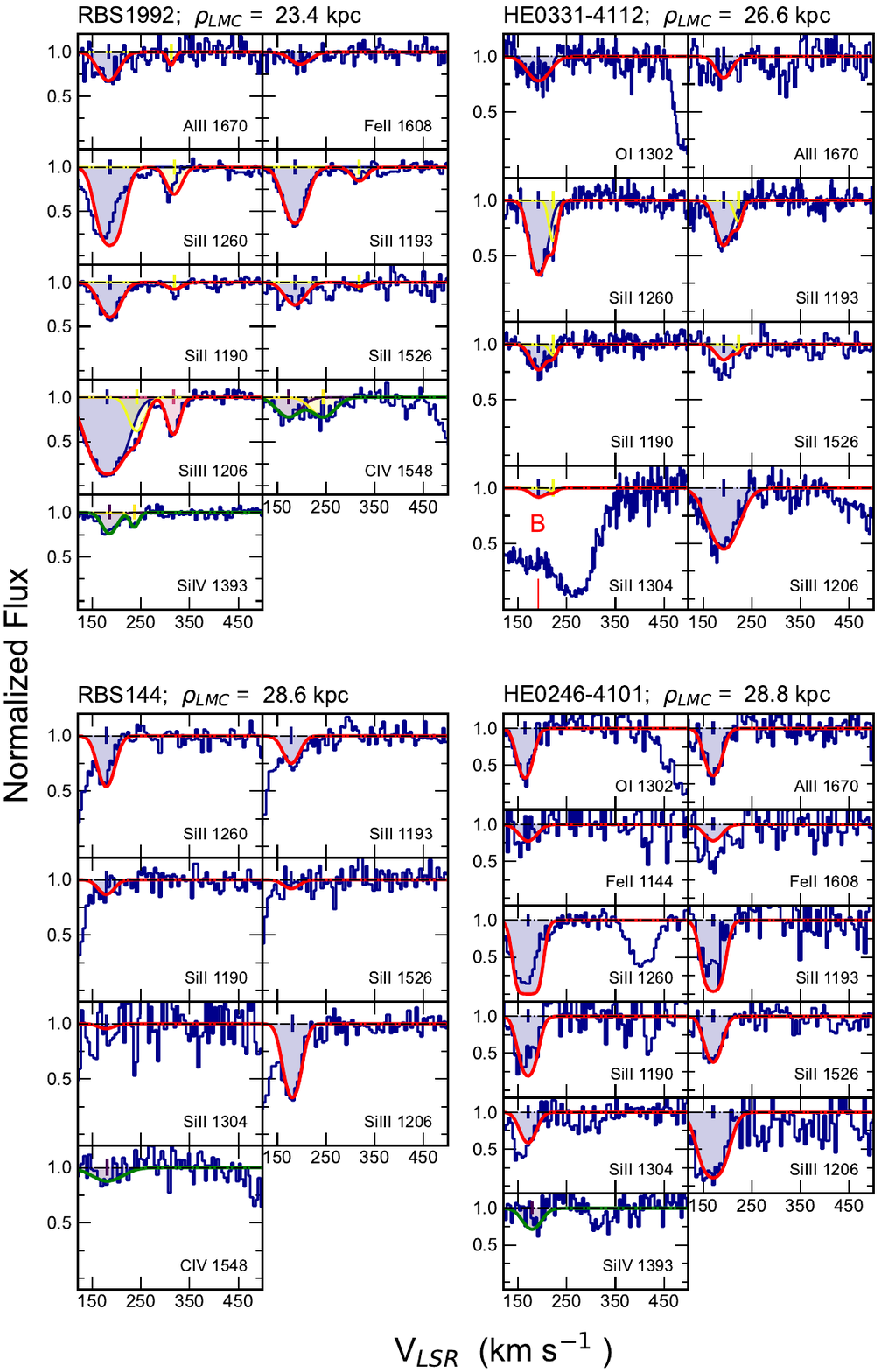}
 \caption{Same as Fig.~\ref{fig:example_spec}.}
 \label{fig:appendix_set5} 
\end{figure*}

\begin{figure*}[!hbt]
 \includegraphics[width=0.95\textwidth]{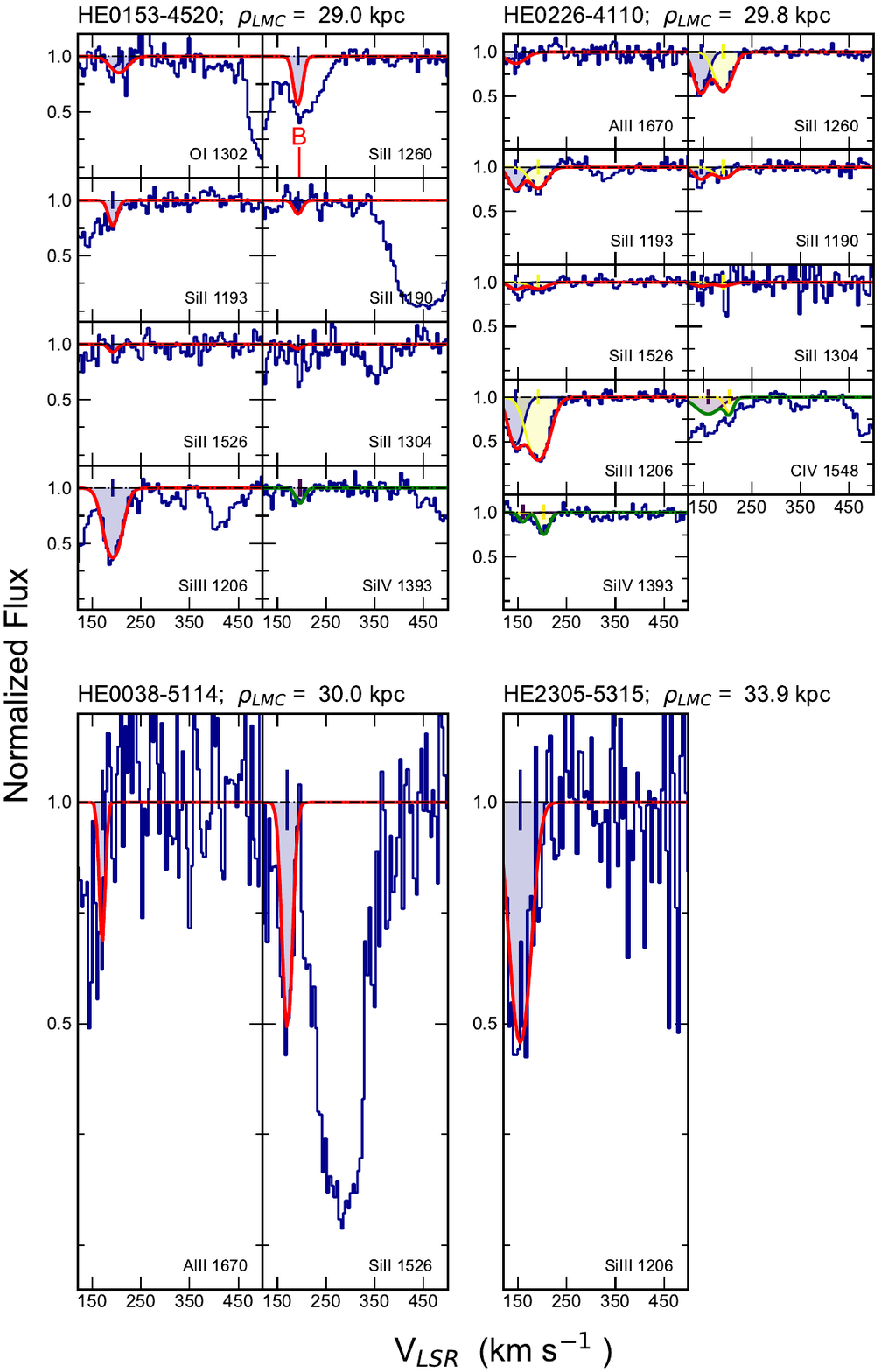}
 \caption{Same as Fig.~\ref{fig:example_spec}.}
 \label{fig:appendix_set6} 
\end{figure*}

\bibliography{Reference}{}

\newcommand{\noop}[1]{}
\begin{thebibliography}{}
\expandafter\ifx\csname natexlab\endcsname\relax\def\natexlab#1{#1}\fi
\providecommand{\url}[1]{\href{#1}{#1}}
\providecommand{\dodoi}[1]{doi:~\href{http://doi.org/#1}{\nolinkurl{#1}}}
\providecommand{\doeprint}[1]{\href{http://ascl.net/#1}{\nolinkurl{http://ascl.net/#1}}}
\providecommand{\doarXiv}[1]{\href{https://arxiv.org/abs/#1}{\nolinkurl{https://arxiv.org/abs/#1}}}

\bibitem[{{Barger} {et~al.}(2016){Barger}, {Lehner}, \& {Howk}}]{Barger2016}
{Barger}, K.~A., {Lehner}, N., \& {Howk}, J.~C. 2016, \apj, 817, 91, \dodoi{10.3847/0004-637X/817/2/91}

\bibitem[{{Barger} {et~al.}(2017){Barger}, {Madsen}, {Fox}, {Wakker}, {Bland-Hawthorn}, {Nidever}, {Haffner}, {Antwi-Danso}, {Hernandez}, {Lehner}, {Hill}, {Curzons}, \& {Tepper-Garc{\'\i}a}}]{Barger2017}
{Barger}, K.~A., {Madsen}, G.~J., {Fox}, A.~J., {et~al.} 2017, \apj, 851, 110, \dodoi{10.3847/1538-4357/aa992a}

\bibitem[{{Besla} {et~al.}(2012){Besla}, {Kallivayalil}, {Hernquist}, {van der Marel}, {Cox}, \& {Kere{\v{s}}}}]{Besla2012}
{Besla}, G., {Kallivayalil}, N., {Hernquist}, L., {et~al.} 2012, \mnras, 421, 2109, \dodoi{10.1111/j.1365-2966.2012.20466.x}

\bibitem[{{Bland-Hawthorn} {et~al.}(2019){Bland-Hawthorn}, {Maloney}, {Sutherland}, {Groves}, {Guglielmo}, {Li}, {Curzons}, {Cecil}, \& {Fox}}]{BlandHawthorn2019}
{Bland-Hawthorn}, J., {Maloney}, P.~R., {Sutherland}, R., {et~al.} 2019, \apj, 886, 45, \dodoi{10.3847/1538-4357/ab44c8}

\bibitem[{{Bordoloi} {et~al.}(2014){Bordoloi}, {Tumlinson}, {Werk}, {Oppenheimer}, {Peeples}, {Prochaska}, {Tripp}, {Katz}, {Dav{\'e}}, {Fox}, {Thom}, {Ford}, {Weinberg}, {Burchett}, \& {Kollmeier}}]{Bordoloi2014}
{Bordoloi}, R., {Tumlinson}, J., {Werk}, J.~K., {et~al.} 2014, \apj, 796, 136, \dodoi{10.1088/0004-637X/796/2/136}

\bibitem[{{Br{\"u}ns} {et~al.}(2005){Br{\"u}ns}, {Kerp}, {Staveley-Smith}, {Mebold}, {Putman}, {Haynes}, {Kalberla}, {Muller}, \& {Filipovic}}]{Bruns2005}
{Br{\"u}ns}, C., {Kerp}, J., {Staveley-Smith}, L., {et~al.} 2005, \aap, 432, 45, \dodoi{10.1051/0004-6361:20040321}

\bibitem[{{Carnall}(2017)}]{Carnall2017}
{Carnall}, A.~C. 2017, arXiv e-prints, arXiv:1705.05165.
\newblock \doarXiv{1705.05165}

\bibitem[{{Carr} {et~al.}(2024){Carr}, {Bryan}, {Garavito-Camargo}, {Besla}, {Setton}, \& {Johnston}}]{Carr2024}
{Carr}, C., {Bryan}, G.~L., {Garavito-Camargo}, N., {et~al.} 2024, arXiv e-prints, arXiv:2408.10358, \dodoi{10.48550/arXiv.2408.10358}

\bibitem[{{Carswell} \& {Webb}(2014)}]{carswell2014}
{Carswell}, R.~F., \& {Webb}, J.~K. 2014, {VPFIT: Voigt profile fitting program}, Astrophysics Source Code Library, record ascl:1408.015

\bibitem[{{de Boer} {et~al.}(1990){de Boer}, {Morras}, \& {Bajaja}}]{Boer1990}
{de Boer}, K.~S., {Morras}, R., \& {Bajaja}, E. 1990, \aap, 233, 523

\bibitem[{{de Boer} \& {Savage}(1980)}]{deBoer1980}
{de Boer}, K.~S., \& {Savage}, B.~D. 1980, \apj, 238, 86, \dodoi{10.1086/157960}

\bibitem[{{Diaz} \& {Bekki}(2011)}]{Diaz2011}
{Diaz}, J., \& {Bekki}, K. 2011, \pasa, 28, 117, \dodoi{10.1071/AS10044}

\bibitem[{{D'Onghia} \& {Fox}(2016)}]{Elena2016}
{D'Onghia}, E., \& {Fox}, A.~J. 2016, \araa, 54, 363, \dodoi{10.1146/annurev-astro-081915-023251}

\bibitem[{{Erkal} {et~al.}(2019){Erkal}, {Belokurov}, {Laporte}, {Koposov}, {Li}, {Grillmair}, {Kallivayalil}, {Price-Whelan}, {Evans}, {Hawkins}, {Hendel}, {Mateu}, {Navarro}, {del Pino}, {Slater}, {Sohn}, \& {Orphan Aspen Treasury Collaboration}}]{Erkal2019}
{Erkal}, D., {Belokurov}, V., {Laporte}, C.~F.~P., {et~al.} 2019, \mnras, 487, 2685, \dodoi{10.1093/mnras/stz1371}

\bibitem[{{Faucher-Gigu{\`e}re} \& {Oh}(2023)}]{Claude2023}
{Faucher-Gigu{\`e}re}, C.-A., \& {Oh}, S.~P. 2023, \araa, 61, 131, \dodoi{10.1146/annurev-astro-052920-125203}

\bibitem[{{Fox} {et~al.}(2013){Fox}, {Richter}, {Wakker}, {Lehner}, {Howk}, {Ben Bekhti}, {Bland-Hawthorn}, \& {Lucas}}]{Fox2013}
{Fox}, A.~J., {Richter}, P., {Wakker}, B.~P., {et~al.} 2013, \apj, 772, 110, \dodoi{10.1088/0004-637X/772/2/110}

\bibitem[{{Fox} {et~al.}(2014){Fox}, {Wakker}, {Barger}, {Hernandez}, {Richter}, {Lehner}, {Bland-Hawthorn}, {Charlton}, {Westmeier}, {Thom}, {Tumlinson}, {Misawa}, {Howk}, {Haffner}, {Ely}, {Rodriguez-Hidalgo}, \& {Kumari}}]{Fox2014}
{Fox}, A.~J., {Wakker}, B.~P., {Barger}, K.~A., {et~al.} 2014, \apj, 787, 147, \dodoi{10.1088/0004-637X/787/2/147}

\bibitem[{{Fujimoto} \& {Sofue}(1977)}]{Fujimoto1977}
{Fujimoto}, M., \& {Sofue}, Y. 1977, \aap, 61, 199

\bibitem[{{Kim} {et~al.}(2024){Kim}, {Zheng}, \& {Putman}}]{Kim2024}
{Kim}, D.~A., {Zheng}, Y., \& {Putman}, M.~E. 2024, \apj, 966, 134, \dodoi{10.3847/1538-4357/ad2def}

\bibitem[{{Krishnarao} {et~al.}(2022){Krishnarao}, {Fox}, {D'Onghia}, {Wakker}, {Cashman}, {Howk}, {Lucchini}, {French}, \& {Lehner}}]{Krishnarao2022}
{Krishnarao}, D., {Fox}, A.~J., {D'Onghia}, E., {et~al.} 2022, \nat, 609, 915, \dodoi{10.1038/s41586-022-05090-5}

\bibitem[{{Kwak} {et~al.}(2015){Kwak}, {Shelton}, \& {Henley}}]{Kwak2015}
{Kwak}, K., {Shelton}, R.~L., \& {Henley}, D.~B. 2015, \apj, 812, 111, \dodoi{10.1088/0004-637X/812/2/111}

\bibitem[{{Lehner} \& {Howk}(2011)}]{Lehner2011}
{Lehner}, N., \& {Howk}, J.~C. 2011, Science, 334, 955, \dodoi{10.1126/science.1209069}

\bibitem[{{Lehner} {et~al.}(2020){Lehner}, {Berek}, {Howk}, {Wakker}, {Tumlinson}, {Jenkins}, {Prochaska}, {Augustin}, {Ji}, {Faucher-Gigu{\`e}re}, {Hafen}, {Peeples}, {Barger}, {Berg}, {Bordoloi}, {Brown}, {Fox}, {Gilbert}, {Guhathakurta}, {Kalirai}, {Lockman}, {O'Meara}, {Pisano}, {Ribaudo}, \& {Werk}}]{Lehner2020}
{Lehner}, N., {Berek}, S.~C., {Howk}, J.~C., {et~al.} 2020, \apj, 900, 9, \dodoi{10.3847/1538-4357/aba49c}

\bibitem[{{Lucchini} {et~al.}(2024){Lucchini}, {D'Onghia}, \& {Fox}}]{Lucchini2024}
{Lucchini}, S., {D'Onghia}, E., \& {Fox}, A.~J. 2024, \apj, 967, 16, \dodoi{10.3847/1538-4357/ad3c3b}

\bibitem[{{Lucchini} {et~al.}(2020){Lucchini}, {D'Onghia}, {Fox}, {Bustard}, {Bland-Hawthorn}, \& {Zweibel}}]{Lucchini2020}
{Lucchini}, S., {D'Onghia}, E., {Fox}, A.~J., {et~al.} 2020, \nat, 585, 203, \dodoi{10.1038/s41586-020-2663-4}

\bibitem[{{Mishra} \& {Muzahid}(2022)}]{Mishra2022}
{Mishra}, S., \& {Muzahid}, S. 2022, \apj, 933, 229, \dodoi{10.3847/1538-4357/ac7155}

\bibitem[{{Moore} \& {Davis}(1994)}]{Moore1994}
{Moore}, B., \& {Davis}, M. 1994, \mnras, 270, 209, \dodoi{10.1093/mnras/270.2.209}

\bibitem[{{Nidever} {et~al.}(2008){Nidever}, {Majewski}, \& {Butler Burton}}]{Nidever2008}
{Nidever}, D.~L., {Majewski}, S.~R., \& {Butler Burton}, W. 2008, \apj, 679, 432, \dodoi{10.1086/587042}

\bibitem[{{Nidever} {et~al.}(2010){Nidever}, {Majewski}, {Butler Burton}, \& {Nigra}}]{Nidever2010}
{Nidever}, D.~L., {Majewski}, S.~R., {Butler Burton}, W., \& {Nigra}, L. 2010, \apj, 723, 1618, \dodoi{10.1088/0004-637X/723/2/1618}

\bibitem[{{Pardy} {et~al.}(2018){Pardy}, {D'Onghia}, \& {Fox}}]{Pardy2018}
{Pardy}, S.~A., {D'Onghia}, E., \& {Fox}, A.~J. 2018, \apj, 857, 101, \dodoi{10.3847/1538-4357/aab95b}

\bibitem[{{Petersen} \& {Pe{\~n}arrubia}(2021)}]{Petersen2021}
{Petersen}, M.~S., \& {Pe{\~n}arrubia}, J. 2021, Nature Astronomy, 5, 251, \dodoi{10.1038/s41550-020-01254-3}

\bibitem[{{Pietrzy{\'n}ski} {et~al.}(2013){Pietrzy{\'n}ski}, {Graczyk}, {Gieren}, {Thompson}, {Pilecki}, {Udalski}, {Soszy{\'n}ski}, {Koz{\l}owski}, {Konorski}, {Suchomska}, {Bono}, {Moroni}, {Villanova}, {Nardetto}, {Bresolin}, {Kudritzki}, {Storm}, {Gallenne}, {Smolec}, {Minniti}, {Kubiak}, {Szyma{\'n}ski}, {Poleski}, {Wyrzykowski}, {Ulaczyk}, {Pietrukowicz}, {G{\'o}rski}, \& {Karczmarek}}]{Pietrzyski2013}
{Pietrzy{\'n}ski}, G., {Graczyk}, D., {Gieren}, W., {et~al.} 2013, \nat, 495, 76, \dodoi{10.1038/nature11878}

\bibitem[{{Prochaska} {et~al.}(2011){Prochaska}, {Weiner}, {Chen}, {Mulchaey}, \& {Cooksey}}]{Prochaska2011}
{Prochaska}, J.~X., {Weiner}, B., {Chen}, H.~W., {Mulchaey}, J., \& {Cooksey}, K. 2011, \apj, 740, 91, \dodoi{10.1088/0004-637X/740/2/91}

\bibitem[{{Putman} {et~al.}(2003{\natexlab{a}}){Putman}, {Bland-Hawthorn}, {Veilleux}, {Gibson}, {Freeman}, \& {Maloney}}]{Putman2003a}
{Putman}, M.~E., {Bland-Hawthorn}, J., {Veilleux}, S., {et~al.} 2003{\natexlab{a}}, \apj, 597, 948, \dodoi{10.1086/378555}

\bibitem[{{Putman} {et~al.}(2003{\natexlab{b}}){Putman}, {Staveley-Smith}, {Freeman}, {Gibson}, \& {Barnes}}]{Putman2003}
{Putman}, M.~E., {Staveley-Smith}, L., {Freeman}, K.~C., {Gibson}, B.~K., \& {Barnes}, D.~G. 2003{\natexlab{b}}, \apj, 586, 170, \dodoi{10.1086/344477}

\bibitem[{{Richter} {et~al.}(2015){Richter}, {de Boer}, {Werner}, \& {Rauch}}]{Richter2015}
{Richter}, P., {de Boer}, K.~S., {Werner}, K., \& {Rauch}, T. 2015, \aap, 584, L6, \dodoi{10.1051/0004-6361/201527451}

\bibitem[{{Richter} {et~al.}(2013){Richter}, {Fox}, {Wakker}, {Lehner}, {Howk}, {Bland-Hawthorn}, {Ben Bekhti}, \& {Fechner}}]{Richter2013}
{Richter}, P., {Fox}, A.~J., {Wakker}, B.~P., {et~al.} 2013, \apj, 772, 111, \dodoi{10.1088/0004-637X/772/2/111}

\bibitem[{{Salem} {et~al.}(2015){Salem}, {Besla}, {Bryan}, {Putman}, {van der Marel}, \& {Tonnesen}}]{Salem2015}
{Salem}, M., {Besla}, G., {Bryan}, G., {et~al.} 2015, \apj, 815, 77, \dodoi{10.1088/0004-637X/815/1/77}

\bibitem[{{Salvatier} {et~al.}(2016){Salvatier}, {Wiecki{\^a}}, \& {Fonnesbeck}}]{Salvatier2016}
{Salvatier}, J., {Wiecki{\^a}}, T.~V., \& {Fonnesbeck}, C. 2016, {PyMC3: Python probabilistic programming framework}, Astrophysics Source Code Library, record ascl:1610.016

\bibitem[{{Setton} {et~al.}(2023){Setton}, {Besla}, {Patel}, {Hummels}, {Zheng}, {Schneider}, \& {Salem}}]{Setton2023}
{Setton}, D.~J., {Besla}, G., {Patel}, E., {et~al.} 2023, \apjl, 959, L11, \dodoi{10.3847/2041-8213/ad0da6}

\bibitem[{{Tumlinson} {et~al.}(2017){Tumlinson}, {Peeples}, \& {Werk}}]{Tumlinson2017}
{Tumlinson}, J., {Peeples}, M.~S., \& {Werk}, J.~K. 2017, \araa, 55, 389, \dodoi{10.1146/annurev-astro-091916-055240}

\bibitem[{{Tumlinson} {et~al.}(2013){Tumlinson}, {Thom}, {Werk}, {Prochaska}, {Tripp}, {Katz}, {Dav{\'e}}, {Oppenheimer}, {Meiring}, {Ford}, {O'Meara}, {Peeples}, {Sembach}, \& {Weinberg}}]{Tumlinson2013}
{Tumlinson}, J., {Thom}, C., {Werk}, J.~K., {et~al.} 2013, \apj, 777, 59, \dodoi{10.1088/0004-637X/777/1/59}

\bibitem[{{Wakker} {et~al.}(1998){Wakker}, {Howk}, {Chu}, {Bomans}, \& {Points}}]{Wakker1998}
{Wakker}, B., {Howk}, J.~C., {Chu}, Y.-H., {Bomans}, D., \& {Points}, S.~D. 1998, \apjl, 499, L87, \dodoi{10.1086/311334}

\bibitem[{{Wakker} \& {van Woerden}(1997)}]{Wakker1997}
{Wakker}, B.~P., \& {van Woerden}, H. 1997, \araa, 35, 217, \dodoi{10.1146/annurev.astro.35.1.217}

\bibitem[{{Wang} {et~al.}(2019){Wang}, {Hammer}, {Yang}, {Ripepi}, {Cioni}, {Puech}, \& {Flores}}]{Wang2019}
{Wang}, J., {Hammer}, F., {Yang}, Y., {et~al.} 2019, \mnras, 486, 5907, \dodoi{10.1093/mnras/stz1274}

\bibitem[{{Watkins} {et~al.}(2024){Watkins}, {van der Marel}, \& {Bennet}}]{Watkins2024}
{Watkins}, L.~L., {van der Marel}, R.~P., \& {Bennet}, P. 2024, arXiv e-prints, arXiv:2401.14458, \dodoi{10.48550/arXiv.2401.14458}

\bibitem[{{Westmeier}(2018)}]{Westmeier2018}
{Westmeier}, T. 2018, \mnras, 474, 289, \dodoi{10.1093/mnras/stx2757}

\bibitem[{{Zheng} {et~al.}(2024{\natexlab{a}}){Zheng}, {Faerman}, {Oppenheimer}, {Putman}, {McQuinn}, {Kirby}, {Burchett}, {Telford}, {Werk}, \& {Kim}}]{Zheng2024}
{Zheng}, Y., {Faerman}, Y., {Oppenheimer}, B.~D., {et~al.} 2024{\natexlab{a}}, \apj, 960, 55, \dodoi{10.3847/1538-4357/acfe6b}

\bibitem[{{Zheng} {et~al.}(2024{\natexlab{b}}){Zheng}, {Tchernyshyov}, {Olsen}, {Choi}, {Bustard}, {Roman-Duval}, {Zhu}, {Di Teodoro}, {Werk}, {Putman}, {McLeod}, {Faerman}, {Simons}, \& {Peek}}]{Zheng2024b}
{Zheng}, Y., {Tchernyshyov}, K., {Olsen}, K., {et~al.} 2024{\natexlab{b}}, arXiv e-prints, arXiv:2402.04313, \dodoi{10.48550/arXiv.2402.04313}

\bibitem[{{Zhu} {et~al.}(2024){Zhu}, {Tonnesen}, {Bryan}, \& {Putman}}]{Zhu2024}
{Zhu}, J., {Tonnesen}, S., {Bryan}, G.~L., \& {Putman}, M.~E. 2024, arXiv e-prints, arXiv:2404.00129, \dodoi{10.48550/arXiv.2404.00129}

\end{thebibliography}
\bibliographystyle{aasjournal}

\end{document}